\pdfoutput=1
\documentclass[10pt,a4paper,twocolumn]{article}

\usepackage[T1]{fontenc}
\usepackage[utf8]{inputenc}
\usepackage{lmodern}
\usepackage[top=22mm,bottom=22mm,left=16mm,right=16mm,columnsep=7mm]{geometry}
\usepackage{microtype}
\usepackage{amsmath,amssymb}
\usepackage{booktabs}
\usepackage{graphicx}
\usepackage{tikz}
\usetikzlibrary{arrows.meta,positioning,fit}
\usepackage{xurl}
\usepackage[round,authoryear]{natbib}
\usepackage[hidelinks]{hyperref}

\hypersetup{
  pdftitle={Multi-Party Authorization of Maintenance Prescriptions on a Permissioned Blockchain},
  pdfauthor={Siyuan Chen, Ebru Turanoglu Bekar, Anders Skoogh}
}

\newcommand{\projectname}{VeriPsM}

\title{\textbf{Multi-Party Authorization of Maintenance Prescriptions on a Permissioned Blockchain}}
\author{Siyuan Chen, Ebru Turanoglu Bekar, and Anders Skoogh\\[3pt]
\small\textit{Department of Mechanical Engineering, Chalmers University of Technology, Gothenburg, Sweden}\\
\small\texttt{\{siyuan.chen, ebrut, anders.skoogh\}@chalmers.se}}
\date{}

\begin{document}
\maketitle

\begin{abstract}
\noindent In industrial prescriptive maintenance, an equipment supplier's model may
recommend an intervention that a plant schedules, a contractor executes, and a
quality or safety function approves. A shared ledger records who submitted the
prescription, yet cannot check whether the agreed model and decision rule
produced it. We require permissioned-blockchain endorsers to verify a succinct
proof of that calculation before releasing a work order. On a four-organization
Hyperledger Fabric testbed using a public engine benchmark, this stopped all
nine contested or mis-operated prescriptions, whereas hash checks accepted four
that either raised simulated maintenance cost or reintroduced late failures.

\end{abstract}

\section{Introduction}
\label{sec:introduction}

Prescriptive maintenance (PsM) turns a condition estimate into a concrete
intervention within resource, risk, and timing constraints
\citep{matyas2017procedural,vanstaden2022predictoptimize}. On industrial
assets maintained under service contracts, that decision crosses company
boundaries: an \emph{equipment supplier} runs its prognostic model to propose
an action, the \emph{plant operator} owns the machine and sensor data and
decides whether to stop production and open a work order, a \emph{maintenance
contractor} carries out the job, and a \emph{quality or safety function} signs
off on the decision rules and compliance records. Digital twins can give these
partners a shared view of asset state \citep{errandonea2020dt,chen2025towards},
yet system integration, data quality, and inter-organizational trust continue
to limit industrial adoption of artificial intelligence (AI)-based maintenance
\citep{chen2025digitaltwins}.

Problems arise when a supplier's recommendation is turned into a binding work
order. A submitted prescription may rest on an outdated condition snapshot, an
unapproved model update, a relaxed risk limit, or a manual edit made after
computation. Once released, any of those errors commits the plant to downtime,
spare parts, contractor hours, or unmitigated failure risk. Even so, the
receiving organizations rarely have the means to rerun the calculation, since
the supplier's model weights and the plant's operating data are confidential.

Permissioned blockchains such as Hyperledger Fabric address part of this
coordination problem through identified membership, endorsement policies, and
an ordered ledger \citep{androulaki2018fabric}, and prior studies have used
them to share rolling-stock maintenance records \citep{abbas2020rollingstock},
coordinate multi-factory maintenance \citep{wang2023collaborative}, and
synchronize distributed manufacturing state \citep{zhang2024trusted}. Yet
standard endorsement and hash commitments only bind who submitted a record and
which asset snapshot, model, or policy identifier was referenced. Whenever an
action is edited in a planning tool, a costlier overhaul is substituted for
convenience, or a risk threshold is relaxed while keeping those references
intact, a hash-based audit trail still accepts the proposal and exposes the
plant to unnecessary shutdowns or missed interventions.

To close this gap, we propose \projectname, a multi-party authorization
workflow in which endorsing organizations verify a succinct cryptographic proof
that the registered model and decision rule produced the submitted
prescription, alongside identity, asset-state, and freshness checks, before a
work order is released. Two questions guide the study:
\textbf{RQ1}: Can multi-party endorsement block altered or unapproved
prescriptions before work-order release without exposing the supplier's model
or the plant's data?
\textbf{RQ2}: How does verifying the prescription calculation affect
maintenance cost, late-event risk, and transaction latency compared with
record-only checks?

We implement and test the workflow on the public PHM North America 2025
aircraft-engine dataset \citep{phm2025challenge}, which supplies multi-sensor
histories and three maintenance event horizons for reproducible evaluation; the
same contractual structure applies wherever an external supplier prescribes
interventions on plant equipment under agreed cost and risk rules. The
contributions are (i) a role-based authorization workflow and deployment
boundary for operator, supplier, contractor, and quality/safety organizations;
(ii) a four-organization Fabric implementation with on-chain kernel parameter
checks; and (iii) an experimental comparison of proof-based, recomputation,
hash-based, and endorsement-only modes across nine prescription alterations and
their simulated maintenance consequences.

\section{Related Work}
\label{sec:related}

\emph{From prognosis to prescription.}
Prognostics and health management (PHM) can estimate degradation and time to
maintenance, but reviews report a persistent gap between model evaluation and
maintenance deployment \citep{nguyen2019phm,sprong2019deployment}. Prescriptive
studies make the decision explicit: \citet{matyas2017procedural} combine
production data and simulation to plan maintenance in manufacturing,
\citet{vanstaden2022predictoptimize} learn a predict-then-optimize policy from
intervention data, and \citet{mao2023complex} add production-planning and
resource constraints in a digital twin. These studies define how an intervention is selected when one organization
holds both the model and the maintenance schedule, leaving open how external
partners can verify that a proposed action came from the agreed calculation.

\emph{Blockchain for maintenance collaboration.}
Permissioned ledgers provide membership, endorsement, and shared state
\citep{androulaki2018fabric}. \citet{abbas2020rollingstock} use a ledger so
that railway stakeholders can trust how rolling stock was maintained,
\citet{wang2023collaborative} coordinate multi-factory production and
collaborative maintenance over a blockchain, and \citet{zhang2024trusted}
couple a digital twin with a blockchain for synchronized control among
distributed manufacturing parties. These approaches improve confidence in
shared records or state updates, while taking the correctness of any off-chain
decision logic on trust.

\emph{Verifiable computation in industrial settings.}
Verifiable computation lets a party check work delegated to an untrusted one
\citep{gennaro2010vc}; succinct proofs \citep{groth2016size} have been applied to
neural predictions \citep{liu2021zkcnn} and, with smart contracts, to model
procurement for industrial cyber-physical systems \citep{burgos2026attestation}.
Maintenance authorization needs more: the evidence must name the asset, its
current state, and the approved policy, and must show that the selected action
beats every declared alternative.

\section{Multi-Party Authorization Workflow}
\label{sec:framework}

\subsection{Roles and workflow}

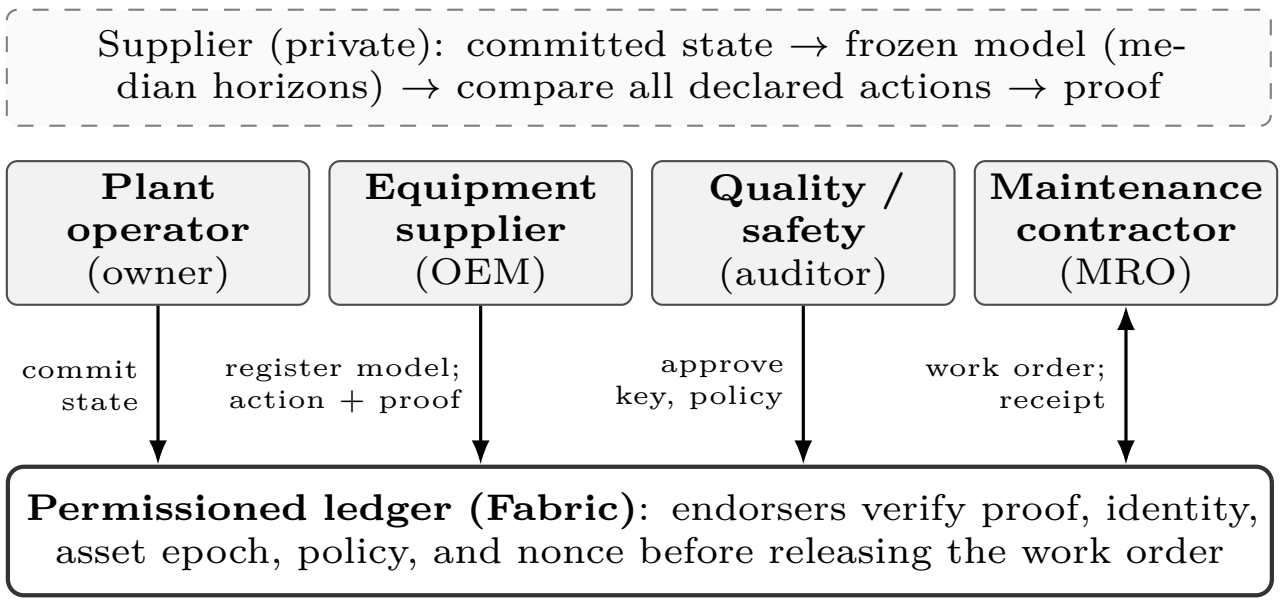
\begin{figure}[t]
\centering
\begin{tikzpicture}[
  org/.style={draw=black!70, rounded corners=2pt, fill=black!5,
    minimum width=2.00cm, minimum height=0.95cm, text width=1.88cm,
    align=center, font=\scriptsize, inner sep=0.06cm},
  led/.style={draw=black!80, thick, rounded corners=3pt, fill=white,
    minimum width=8.36cm, minimum height=0.85cm, text width=8.12cm,
    align=center, font=\scriptsize, inner sep=0.12cm},
  flow/.style={-{Latex[length=1.6mm]}, line width=0.6pt},
  biflow/.style={{Latex[length=1.6mm]}-{Latex[length=1.6mm]}, line width=0.6pt},
  node distance=0.12cm]
\node[org] (op) {\textbf{Plant}\\\textbf{operator} (owner)};
\node[org, right=of op] (oem) {\textbf{Equipment}\\\textbf{supplier} (OEM)};
\node[org, right=of oem] (aud) {\textbf{Quality /}\\\textbf{safety} (auditor)};
\node[org, right=of aud] (mro) {\textbf{Maintenance}\\\textbf{contractor} (MRO)};
\node[led, below=1.05cm of op.south west, anchor=north west] (ledger)
  {\textbf{Permissioned ledger (Fabric)}: endorsers verify proof, identity,
   asset epoch, policy, and nonce before releasing the work order};
\node[draw=black!50, dashed, rounded corners=2pt, fill=black!2,
  minimum width=8.36cm, font=\scriptsize, text width=8.12cm, align=center,
  inner sep=0.12cm, above=0.22cm of op.north west, anchor=south west] (calc)
  {Supplier (private): committed state $\rightarrow$ frozen model (median
   horizons) $\rightarrow$ compare all declared actions $\rightarrow$ proof};
\draw[flow] (op.south) -- node[left, font=\tiny, align=right]{commit\\state} (op.south |- ledger.north);
\draw[flow] (oem.south) -- node[left, font=\tiny, align=right]{register model;\\action + proof} (oem.south |- ledger.north);
\draw[flow] (aud.south) -- node[left, font=\tiny, align=right]{approve\\key, policy} (aud.south |- ledger.north);
\draw[biflow] (mro.south |- ledger.north) -- node[left, font=\tiny, align=right]{work order;\\receipt} (mro.south);
\end{tikzpicture}
\caption{Multi-party authorization of a maintenance prescription. The supplier
runs the model privately, and endorsing peers verify the proof and public
bindings without accessing model weights or sensor data.}
\label{fig:framework}
\end{figure}

Fig.~\ref{fig:framework} shows the four organizations, which share one Fabric
channel and appear in the testbed as Operator, original equipment manufacturer
(OEM), maintenance, repair, and overhaul (MRO), and Auditor. The \emph{plant
operator} (asset owner) commits the current asset state. The \emph{equipment
supplier} (OEM) registers its frozen prognostic model and the decision policy
and submits prescriptions. The \emph{quality/safety function} (auditor)
approves the policy, its risk limit and optimality tolerance, and the
verification key before any prescription can be accepted. The \emph{maintenance
contractor} (MRO) receives authorized work orders and records execution
receipts. Each step is a separate ledger transaction, so responsibility for
every hand-over is attributable.

A submitted prescription is accepted only if its public fields match the
active registration and current asset state, its nonce is unused, and its proof
verifies on every endorsing peer. Cryptographic verification thus operates
alongside channel identity, kernel approval, and contractor execution receipts.

\subsection{Prescription statement and decision rule}

Each prescription carries a public statement
\begin{equation}
S = (a, e, c_s, m, p, u, \tau, r, \varepsilon, n, q),
\label{eq:statement}
\end{equation}
naming the asset, state epoch, state commitment, model, policy, action, action
time, risk limit, tolerance, nonce, and proof type. The private part holds 29
integer model inputs, resource capacity, a commitment salt, and the selected
candidate. The proof shows that the registered model, applied to the committed
inputs, yields median event horizons (q50) for three targets (water wash,
compressor shop visit, turbine shop visit), and that the selected action is
cost-optimal among the declared candidates.

The registered policy declares four candidates representing generic industrial
interventions: defer by 500 cycles (routine deferral), water wash now (online
servicing), compressor shop visit now (subsystem overhaul), or turbine shop
visit now (major overhaul). For each target, acting no later than q50 carries
50,000 parts per million (ppm) risk and acting later 900,000 ppm; candidate
risk is the maximum over targets, weighted by 25 cost units per 1,000 ppm. A
candidate is admissible if its resource demand fits plant capacity and its risk
does not exceed the auditor-approved limit $r$ (550,000 ppm). Each integer
objective $J_k$ combines direct and downtime cost with early-maintenance and
risk terms, and an admissible selected candidate $k^\star$ must satisfy
\begin{equation}
J_{k^\star}\leq J_k+\varepsilon\quad\text{for every admissible }k
\label{eq:optimality}
\end{equation}
(with $\varepsilon=0$). Predicted horizons outside $[0,32767]$ cycles fail
range verification without clamping.

\subsection{One decision cycle}

After a production shift, the plant operator commits the latest condition
snapshot of a machine, advancing its state epoch. The supplier's model, running
in the supplier's or the plant's environment, estimates the remaining horizons
and the planner compares the four candidates; the supplier then submits the
selected action with its proof. Endorsing peers from the four organizations
check that the statement names the current epoch and the approved model,
policy, and risk parameters, and that the proof verifies. When those checks
pass, the ledger records an authorized work order for the contractor, who
writes back an execution receipt upon completion; the next state commitment
opens the following epoch. Deferrals pass through the same endorsement path as
interventions, leaving an auditable record when production continues without a
stop.

\subsection{Deployment boundary}
\label{sec:boundary}

Governance and membership are fixed before routine operation. The action set,
cost weights, risk limit, capacity bounds, and frozen model manifest are
registered on-chain and approved by the quality/safety organization; changing
model weights or cost parameters requires registering and approving a new
kernel identifier and verification key. Each organization operates its own peer
and certificate authority, and endorsement requires the verifying peers to
concur.

Raw sensor histories, model weights, and private feature vectors remain at the
plant or supplier, while the ledger stores only commitments, identifiers, the
selected action, and the 164-byte proof. When a proposal fails a binding or
proof check, endorsement aborts before ordering, leaving the asset record
unchanged and returning the rejection reason to the submitter; urgent safety
stops continue to use the plant's normal manual work-order procedure.

In a plant deployment, an authorized-\hspace{0pt}prescription event on the
ledger would trigger work-order creation in the computerized maintenance
management system (CMMS), with the execution receipt written back at order
closure; our testbed implements only the ledger side of that interface.
Correctness still depends on trustworthy sensor measurements, an appropriate
registered model and policy, and sound key and proof-setup management, and
public asset tags or action codes on the channel can reveal maintenance cadence
to channel members.

\section{Benchmark Case and Implementation}
\label{sec:casestudy}

\emph{Data and prognostic model.} The public PHM North America 2025 engine
data \citep{phm2025challenge} contain 59,702 labelled records from four engines;
testing and validation files for eight further engines lack local outcome
labels and are excluded from fitting and accuracy reporting. The model uses 14
shared sensor fields with leave-one-engine-out folds. A current-sensor
LightGBM teacher predicts the three maintenance horizons, and a one-hidden-layer
integer student reproduces the decision-relevant forecasts with frozen weights,
scales, and ranges. Because only four engines carry ground-truth labels, the prognostic fit serves
solely to supply realistic inputs for testing the authorization workflow.

\emph{Proof and network.} The proof uses gnark v0.12.0 \citep{gnark} with
Groth16 \citep{groth2016size}; it evaluates the student, converts its outputs
exactly to q50 values, and recomputes all four candidate objectives. Setup keys
are local test material, and endorsing chaincode containers cache the approved
verification key by digest. The Hyperledger Fabric 2.5.16 testbed has four peer
organizations, one peer each, three Raft orderers, and one channel. Four
authorization modes share the same chaincode interface: \emph{proof-based}
(proof plus governance checks), \emph{recompute} (endorsers rerun the model and
rule on disclosed inputs), \emph{hash-only} (commitment hashes and freshness),
and \emph{endorsement-only} (signatures only). Each mode receives one valid
prescription and nine altered ones, and the asset record is queried after every
rejection.

\emph{Maintenance consequences.} A semi-synthetic simulator compares deferral,
a median-threshold heuristic, and the proved rule under common random numbers,
crossing intervention cost, reset effectiveness, and risk cost at three levels
(27 settings, 100 paired seeds, 20 decisions per seed). It reports cost in
normalized units and late maintenance events, and projects accepted alterations
under three tampering regimes: action inflation (10\% of deferrals replaced by
major overhauls), intervention suppression (80\% of prescribed interventions
deferred), and bidirectional flips (10\%).

\section{Results}
\label{sec:results}

\subsection{Rejection of altered prescriptions}

Table~\ref{tab:attacks} lists the nine altered prescriptions as the maintenance
situations they represent. All modes accepted the valid prescription.
Proof-based authorization rejected every alteration, and rejected proposals
left the asset record unchanged; in particular, an unregistered risk limit was
blocked by checking the auditor-approved kernel parameters on-chain even before
pairing verification. Hash checks stopped the five situations that
change \emph{which asset, state, model, or policy} is referenced, but accepted
the four that change \emph{the decision itself} while the references stay
intact. Endorsement alone accepted everything. Recomputation on disclosed
inputs rejected the same decision changes and accepted only a corrupted proof
attached to the correct action.

\begin{table*}[t]
\centering
\caption{Acceptance of nine contested or mis-operated prescriptions on the
four-organization Fabric testbed ($\times$ = rejected, \checkmark{} =
accepted and released as a work order). All modes accepted the valid control.}
\label{tab:attacks}
\footnotesize
\setlength{\tabcolsep}{4pt}
\begin{tabular}{llcccc}
\toprule
Alteration & Plant-level situation & Proof & Recompute & Hash & Endorse \\
\midrule
State substitution      & Prescription computed on an edited or outdated asset condition & $\times$ & $\times$ & $\times$ & \checkmark \\
Model substitution      & Supplier deploys an unapproved model update & $\times$ & $\times$ & $\times$ & \checkmark \\
Policy substitution     & Costs or action set changed without quality/safety approval & $\times$ & $\times$ & $\times$ & \checkmark \\
Stale epoch             & Old prescription resubmitted after the asset state changed & $\times$ & $\times$ & $\times$ & \checkmark \\
Cross-asset reuse       & Prescription for one machine reused for another & $\times$ & $\times$ & $\times$ & \checkmark \\
Action tampering        & Action edited after computation (e.g., in a planning tool) & $\times$ & $\times$ & \checkmark & \checkmark \\
Forged evidence         & Supporting evidence fabricated or corrupted & $\times$ & \checkmark & \checkmark & \checkmark \\
Non-optimal action swap & A costlier or more convenient action chosen instead of the optimum & $\times$ & $\times$ & \checkmark & \checkmark \\
Unregistered risk limit & Risk threshold relaxed or tightened outside the approved policy & $\times$ & $\times$ & \checkmark & \checkmark \\
\midrule
\multicolumn{2}{l}{Altered prescriptions released as work orders} & 0/9 & 1/9 & 4/9 & 9/9 \\
\bottomrule
\end{tabular}
\end{table*}

\subsection{Maintenance consequences of accepted alterations}

In the paired simulation (central setting), the industrial consequence of the
four alterations accepted by hash checks depended on the tampering direction.
Under action inflation, unscheduled major overhauls raised mean paired cost by
44,960 to 55,066 normalized units over the clean policy's 31,845 (all 95\%
bootstrap intervals above $+36{,}458$). Under intervention suppression,
delaying prescribed maintenance lowered apparent short-term expenditure by
19,651 to 21,528 units while raising mean late maintenance events from 0.00 to
between 0.70 and 0.83 per seed, affecting 65 to 75 of the 100 seeds.
Bidirectional flips raised mean paired cost by 43,490 to 51,697 units.
Endorsement-only authorization released all nine alterations with comparable
cost surges (44,960 to 55,066) or late-event exposure (0.70 to 0.83 across 65
to 77 seeds), whereas proof-based and recomputation modes kept action flips and
paired cost and risk deltas at zero.

\subsection{Policy cost and late-event comparison}

Across all 2,700 scenario-seed runs (54,000 decisions), the proved rule
produced zero horizon-domain rejections and zero late maintenance events,
selecting 34,155 deferrals, 18,828 online washes, and 1,017 major overhauls. In
the central setting (100 paired seeds), always deferring incurred 7,650 mean
cost and 7.65 downtime cycles but suffered 1.53 mean late events (all 100 seeds
affected). The unproved median-threshold heuristic eliminated late events at a
mean cost of 45,105 and 28.29 downtime cycles. The proved rule also eliminated
all late events (0.00) while lowering mean cost by 29\% to 31,845 (95\%
interval $[28{,}396, 35{,}457]$) and downtime to 21.63 cycles (63.5\%
deferrals, 34.7\% water washes, 1.9\% turbine visits), because its integer
objective balances early-maintenance waste against threshold risk.

\subsection{Proof and ledger performance}

Independent Python, Go, and circuit implementations produced identical integer
accumulators, and exact q50 conversion agreed on all 179,106 labelled
comparisons across 59,702 records with zero domain rejections (maximum q50:
15,287 cycles). Across the benchmark snapshots, the proved rule selects all
four candidates (19,913 deferrals, 36,607 washes, 2 compressor visits, 3,180
turbine visits), and valid Groth16 proofs verified for real snapshots of every
action and domain boundary while rejecting all 15 mutation classes
(Table~\ref{tab:proof}).

\begin{table}[t]
\centering
\caption{Integrated proof (medians of five local runs; verification averaged
over 100 calls per run) and grouped prognosis score (macro score over four
held-out engines; lower is better).}
\label{tab:proof}
\footnotesize
\setlength{\tabcolsep}{4pt}
\begin{tabular}{lr}
\toprule
Measure & Result \\
\midrule
Constraints (total / planner) & 29,048 / 11,010 \\
Proof / key size & 164 B / 780 B \\
Prove / verify time & 52.71 ms / 0.381 ms \\
Conversion mismatches & 0 / 179,106 \\
Domain rejections & 0 / 59,702 \\
Proved actions / mutations & 4 of 4 / 15 of 15 \\
\midrule
Constant ref.\ / integer student & 136.93 / 85.29 \\
\bottomrule
\end{tabular}
\end{table}

\begin{figure}[t]
\centering
\includegraphics[width=8.4cm]{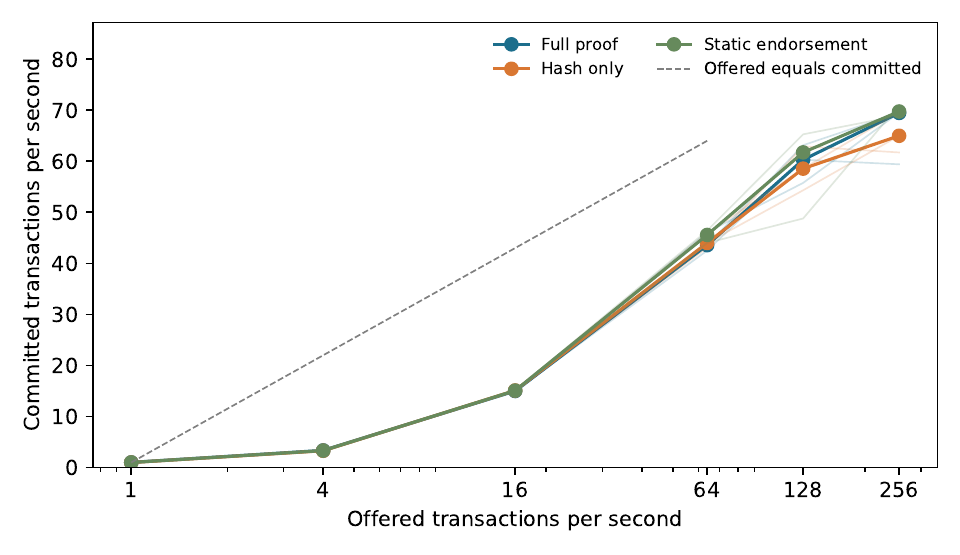}
\caption{Committed versus offered throughput in short bursts. Solid lines are
medians over three fresh ledgers; the dashed line marks equality up to 64
transactions per second.}
\label{fig:throughput}
\end{figure}

Across three fresh ledgers all 1,620 measured transactions committed. At one
transaction per second the median round trip was 1,344 ms for proof-based,
1,343 ms for hash-only, and 1,343 ms for endorsement-only authorization;
verification inside the endorsing peers took a median of 0.703 ms (peer-restart
medians 0.608--0.782 ms), about 0.05\% of the round trip
(Fig.~\ref{fig:throughput}). With verification-key caching, across-restart
median committed throughput at 256 offered transactions per second reached
69.5, 65.0, and 69.7 transactions per second for proof-based, hash-only, and
endorsement-only modes (median round trips 341, 344, and 334 ms), and the
per-restart throughput ranges overlapped across all three modes.

\section{Discussion}
\label{sec:discussion}

\subsection{Comparison with prior studies}

Table~\ref{tab:attacks} clarifies how calculation verification extends prior
blockchain applications in maintenance and manufacturing. Platforms for
rolling-stock maintenance records \citep{abbas2020rollingstock}, multi-factory
collaborative maintenance \citep{wang2023collaborative}, and digital-twin state
synchronization \citep{zhang2024trusted} use permissioned ledgers to
authenticate participants and anchor shared state hashes, while treating
off-chain optimization outputs as given. That boundary protects provenance of
the inputs yet leaves the decision step itself exposed during work-order
hand-overs, where manual action edits, convenient overhaul substitutions, or
relaxed risk limits can pass hash checks under valid signatures. Regarding RQ1
and RQ2, enforcing proof verification at endorsement closes this hand-over gap
without disclosing model weights or plant sensor histories, protecting both the
operator against inflated overhaul spend and the safety function against
deferred interventions whose short-term cost savings mask rising failure risk.
Because pairing verification is orders of magnitude shorter than Fabric
endorsement and ordering, adding calculation checks strengthens work-order
governance without degrading consensus latency or burst throughput.

The workflow also connects single-enterprise prescriptive maintenance with
verifiable machine learning. Existing prescriptive methods couple prognosis
with cost- and resource-aware action selection
\citep{matyas2017procedural,vanstaden2022predictoptimize,mao2023complex} under
the assumption that one organization controls both model inference and work
execution; in cross-company service contracts, confidentiality and misaligned
incentives slow practical adoption \citep{sprong2019deployment,chen2025digitaltwins}.
Meanwhile, succinct proofs have been demonstrated for standalone neural
predictions \citep{liu2021zkcnn} and model-procurement benchmarks
\citep{burgos2026attestation}, without binding asset-epoch freshness,
auditor-approved risk bounds, and candidate optimality inside multi-party
endorsement \citep{androulaki2018fabric}. Where partners can share weights and
sensor snapshots inside one corporate boundary, on-chain recomputation provides
the same protection against decision tampering more simply; cryptographic
proofs become worthwhile when supplier models or plant data are proprietary,
and additionally prevent corrupted evidence from entering the audit archive.

Operationally, requiring a proof at endorsement moves policy alignment into
contract setup: operator, OEM, contractor, and auditor agree up front on
candidate actions, cost weights, and risk limits, settling rule compliance
before work-order release. Subsequent disagreements over a maintenance outcome
then center on whether the agreed parameters reflect plant economics and risk
appetite, since adherence to the registered rule is established at endorsement.
Because the chaincode and circuit depend only on a committed feature vector, a
frozen integer horizon model, and priced candidates, the same structure
transfers to machine-tool spindles, process compressors, and robotic cells
maintained under OEM service agreements
\citep{errandonea2020dt,chen2025digitaltwins}, where larger models or action
sets increase off-chain proving effort while proof size and peer verification
cost remain constant on-chain.

\subsection{Limitations and future work}

First, the public engine benchmark includes ground-truth labels for only a
small cohort \citep{phm2025challenge}, making the prognostic fit exploratory,
and the maintenance consequences are derived from a stylized renewal simulator
under three tampering regimes. Second, although cost-and-risk calibration
enabled the proved policy to balance online servicing, overhauls, and deferrals
without late failures, the circuit still evaluates an isolated asset at a
single decision epoch under a binary risk step, whereas plant maintenance
planning couples multiple machines through shared crews, spare-part
inventories, and opportunistic shutdown windows
\citep{mao2023complex,wang2023collaborative}. Future work will extend
(\ref{eq:optimality}) to multi-asset window scheduling with piecewise-linear
hazard approximations and validate both the prognostic and cost parameters on
industrial plant datasets.

Third, the testbed runs one peer per organization on a single host, uses local
test certificates and Groth16 setup keys \citep{groth2016size}, evaluates
short transaction bursts instead of sustained multi-host traffic, and stops at
the ledger boundary without a live CMMS connector, while covering a finite set
of contract-boundary faults. Corresponding future work will test multi-host
endorsement under sustained plant workloads, replace local setup keys with
ceremony-managed or transparent proof backends, and close the loop from ledger
work-order events and contractor receipts back to plant CMMS records and
digital-twin state commitments \citep{errandonea2020dt,chen2025towards}.

\section{Conclusion}
\label{sec:conclusion}

Permissioned blockchains can coordinate equipment suppliers, plant operators,
contractors, and quality/safety functions in prescriptive maintenance, yet
standard endorsement checks only signatures and record hashes. \projectname{}
requires endorsing peers to verify a succinct proof of the prescription
calculation alongside approved policy bounds before releasing a work order. On
a four-organization Fabric testbed with a public engine benchmark, a calibrated
proved policy eliminated simulated late events at 29\% lower cost than a
threshold heuristic and rejected all nine contested or mis-operated
prescriptions, compared with four accepted under hash checks and nine under
endorsement alone, at 0.703 ms median peer verification time and with matching
short-burst throughput across modes.

\section*{Acknowledgements}
The work was carried out within Chalmers' Area of Advanced Production whose
support is greatly acknowledged. The computation was enabled by resources
provided by the National Academic Infrastructure for Supercomputing in Sweden
(NAISS), partially funded by the Swedish Research Council through grant
agreement no.~2022-06725.

\bibliographystyle{plainnat}
\bibliography{references}

\end{document}